\documentclass[12 pt]{amsart}
\usepackage{mathrsfs}
\usepackage{graphicx}
\usepackage{epstopdf}

 \theoremstyle{definition}
 
 \theoremstyle{remark}

 \numberwithin{equation}{section}

\begin{document}
\title[Gaussian coherence breaking channel]{Characterizing the Gaussian coherence breaking channel and its property with assistant entanglement inputs}

\author{Kan He}
\address[Kan He]{College of
Mathematics, Institute of Mathematics, Taiyuan University of
Technology, Taiyuan,
 030024, P. R. China} \email[K. He]{ hk19830310@163.com}

\author{Jinchuan Hou}
\address[Jinchuan Hou]{College of
Mathematics, Institute of Mathematics, Taiyuan University of
Technology, Taiyuan,
 030024, P. R. China} \email[J. Hou]{houjinchuan@tyut.edu.cn}

 \maketitle

 \begin{abstract}
 We give a characterization of arbitrary $n$-mode Gaussian
coherence breaking channels (GCBCs)  and show that the tensor
product of a GCBC with an arbitrary  Gaussian channel maps all input
states into product states. The inclusion relations between the sets
of GCBCs, Gaussian positive partial transpose channels (GPPTCs),
entanglement breaking channels (GEBCs) and Gaussian
classical-quantum channels (GCQCs) are displayed.
\end{abstract}

\section{Introduction}

Quantum coherence is one of important topics of the quantum theory
and has been developed as an important quantum resource in recent
years (Ref. \cite{SAP}). Furthermore, it had been linked to quantum
entanglement in many quantum phenomenon and plays an important role
in fields of quantum biology and quantum thermodynamics
(\cite{SS}-\cite{SBP}).

In finite-dimensional systems, the theory of (quantifying) quantum
coherence has been undertaken by many authors
(\cite{Gl}-\cite{SFL}). Quantum operations relative to coherence
have become important approaches and resources, such as incoherent
operations. Recently, coherence breaking channels for the
finite-dimensional case, as a subclass of inherent operations, are
defined and characterized in \cite{BSSW}. As we know, continuous
variable (CV) quantum systems are fundamental important from
theoretical and experimental views. In particular, Gaussian states
can be produced and managed experimentally. Several researchers
focused on the theory of (quantifying) quantum coherence of Gaussian
states (\cite{BCP}-\cite{Xu}). Furthermore, Gaussian incoherent
channels also be characterized in \cite{Xu}. The purpose of this
paper is to characterize Gaussian coherence-breaking channels and
discuss some relative topics.

The paper is organized as follows. In Section 2, we give a complete
characterization of Gaussian coherence-breaking channels.
Furthermore, we show the inclusion relation among the sets of
Gaussian positive partial transpose channels (GPPTCs), entanglement
breaking channels (GEBCs) and Gaussian classical-quantum channels
(GCQCs). In Section 3, we discuss the properties of the tensor
product of a GCBC with an arbitrary Gaussian channel with
entanglement inputs, and show  that the tensor product of a GCBC and
an arbitrary   Gaussian channel maps all input states into product
ones. Its connection to the channel capacity is also discussed.

\section{Characterizing GCBCs}

We first recall some necessary notions.

 Let $H$ be the complex separable
infinite-dimensional Hilbert space, $\mathcal T(H)$ the trace class
operators on $H$. For the fixed reference basis $\{|i\rangle\}$, a
state $\rho$ is incoherent if $\rho=\sum_{i} \lambda_i |i\rangle
\langle i|$; otherwise, it is coherent. Denote by $\mathcal I_C$ the
set of all incoherent states and $\mathcal I_C^G$ the set of all
incoherent Gaussian states.

{\bf Definition 2.1} A channel $\Phi: \mathcal T(H_A)\rightarrow
\mathcal T(H_B)$ is called a Gaussian coherence-breaking channel
(GCBC) if $\Phi(\rho)$ is always incoherent for any Gaussian input
state $\rho\in \mathcal T(H_A)$.


The aim of this section is to give a  characterization of GCBCs.

Recall that an $n$-mode Gaussian state $\rho$ is represented by
$$\rho=\frac{1}{(2\pi)^n }\int {\rm d}^{2n}z\  {\rm exp}(-\frac{1}{4} z^T \nu z + i d^T z)W(-z), \eqno(2.1)$$
where $W(z)$ is the Weyl operator, $\nu$ is the covariance matrix
(CM) of $\rho$, which is a $2n \times 2n$ real symmetric matrix, and
$d$ is the displacement vector, which is a $2n$-dimensional real
vector. So a Gaussian state is described by its covariance matrix
and displacement and thus  we write $\rho=\rho[\nu, d]$.

If $\Phi$ is an $n$-mode (Bosonic) Gaussian channel, then for any
input $\rho=\rho[\nu,d]$, the output $\Phi(\rho)$ has the following
covariance matrix and displacement:
$$ K\nu K^T+M,\ Kd+\bar{d}, $$
where $\bar{d}$ is a $2n$ dimensional real vector, $K, M$ are
$2n\times 2n$ real matrices satisfying $M\geq \pm \frac{i}{2}(\Delta
- K\Delta K^T)$, $K^T$ is the transpose of $K$ and
$$\Delta=\oplus^n
 \left(\begin{array} {ccccccccc} 0 &
 1 \\
 -1 & 0\\
 \end{array}\right).$$
  So we write a Gaussian channel  as $\Phi=\Phi(K, M, \bar{d})$.

The form of the multi-mode incoherent Gaussian operations were
described implicitly  in \cite{Xu} without any proof. However, to
obtain a characterization of GCBCs, we need formulate the multi-mode
incoherent Gaussian operations exactly.  We do this in the following
lemma.

Note that, for any $2\times 2$ real orthogonal matrix ${\mathcal
O}$, we have ${\mathcal O}\left(\begin{array}{cc} 0 & 1 \\ -1 &
0\end{array}\right){\mathcal O}^T= \left(\begin{array}{cc} 0 & 1
\\ -1 & 0\end{array}\right)$ or $-\left(\begin{array}{cc} 0 & 1 \\ -1 &
0\end{array}\right) $ according to $\mathcal O$ is symplectic or
not.

{\bf Lemma 2.1} {\it An $n$-mode Gaussian channel $\Phi=\Phi(K, M,
\bar d)$   satisfies $\Phi(\mathcal I_C^G)\subseteq \mathcal I_C^G$
if and only if $\bar{d}=0$ and there exists a permutation $\pi$ on
$\{1,2, ..., n\}$ such that \if false $$\Phi=\otimes_{i=1}^n
\Phi_{i\mapsto \pi(i)},\eqno(2.1)$$ where, for each $i$,
$\Phi_{i\mapsto \pi(i)}$ is a Gaussian channel from the $i$th mode
to the $\pi(i)$th mode. Moreover, let  then \fi
$$K=(P_\pi\otimes I_2) (\oplus_{i=1}^n t_i \mathcal O_i), \eqno(2.2)$$
 $$M=\oplus_{i=1}^n  \lambda_{\pi(i)}I_2, \eqno(2.3)$$ where $P_\pi$ is the $n\times n$ permutation matrix
corresponding to $\pi$, $\mathcal O_i$s are some $2\times 2$
orthogonal matrices and   $\lambda_i\geq \frac{1}{2}|t_i^2- 1|$
whenever ${\mathcal O}_i$ is symplectic or   $\lambda_i\geq
\frac{1}{2}|t_i^2+ 1|$ otherwise.}

If $K$ and $M$ have respectively the form in Eq.(2.2) and Eq.(2.3),
let $\Phi_{i\to \pi(i)}=\Phi(K_{\pi(i)}, \lambda_{\pi(i)}I_2, 0)$ be
the Gaussian channel from the $i$th mode into the $\pi(i)$th mode
determined by $K_{\pi(i)}$, $\lambda_{\pi(i)}I_2$ and $\bar{d}=0$.
Then $\Phi(K,M,\bar{d})=\otimes_{i=1}^n \Phi_{i\mapsto \pi(i)}$ and
every $\Phi_{i\to\pi(i)}$ is an incoherent Gaussian channel between
$i$th and $\pi(i)$th one-mode systems by \cite[Theorem 2]{Xu}. Then
the following corollary is obvious.

{\bf Corollary 2.2} {\it An $n$-mode Gaussian channel $\Phi$
satisfies $\Phi(\mathcal I_C^G)\subseteq \mathcal I_C^G$ if and only
if there exists a permutation $\pi$ on $\{1,2, ..., n\}$ and
one-mode incoherent Gaussian channel $\Phi_{i\mapsto \pi(i)}$ from
$i$th mode into $\pi(i)$th mode for each $i$ such that
 $$\Phi=\otimes_{i=1}^n \Phi_{i\mapsto \pi(i)}.$$
}

{\bf Proof of Lemma 2.1.} Noting that a single-mode Gaussian state
is incoherent (thermal) if and only if it has a diagonal covariance
matrix, zero displacement and squeezing; and every $n$-mode Gaussian
incoherent state is a tensor product of single-mode Gaussian
incoherent states (\cite{BNP}, \cite{Xu}). Then, for any $n$-mode
Gaussian channel $\Phi=\Phi(K, M, \bar{d})$, if $\Phi(\mathcal
I_C^G)\subseteq \mathcal I_C^G$, it is obvious that $\bar d=0$.

Write $K=(A_{ij})_{n\times n}$ and $M=(M_{ij})_{n\times n}$ with
 $A_{ij}, M_{ij}$ are $2\times 2$ real matrices. Note that an
$n$-mode incoherent Gaussian state $\rho$ has the covariance matrix
$$V_\rho={\rm diag}( r_1 I_2, r_2 I_2, ... , r_n I_2).$$
 Since $\Phi$
maps arbitrary incoherent states to incoherent ones, we have that,
for each  set $\{r_i\}_{i=1}^n$ corresponding to an input incoherent
Gaussian state, there exist a set $\{s_i\}_{i=1}^n$ corresponding to
the output state such that
$$\begin{array}{llllll} & KV_\rho K^T +M = (A_{ij})_{n\times
n}(\oplus_{i=1}^n r_i I_2) (A_{ij})_{n\times n}^T+M \\ & = \left(
  \begin{array}{llllll}
   \sum_{j=1}^n r_jA_{1j}A_{1j}^T &  \sum_{j=1}^n r_jA_{1j}A_{2j}^T &  \ldots & \sum_{j=1}^n r_jA_{1j}A_{nj}^T\\
    \sum_{j=1}^n r_jA_{2j}A_{1j}^T &  \sum_{j=1}^n r_jA_{2j}A_{2j}^T& \ldots & \sum_{j=1}^n r_jA_{2j}A_{nj}\\
\vdots & \vdots & \ddots &\vdots \\
    \sum_{j=1}^n r_jr_jA_{nj}A_{1j}^T &  \sum_{j=1}^n r_j A_{nj}A_{2j}^T & \ldots & \sum_{j=1}^n r_jA_{nj}A_{nj}^T\\
\end{array}\right)+M \\ &= \oplus_{i=1}^n s_i I_2\end{array}$$
It follows  that for each pair $(k,l)$,
$$\sum_{j=1}^n r_jA_{kj}A_{kj}^T+M_{kk}=s_kI_2;$$$$  \sum_{j=1}^n
r_jA_{kj}A_{lj}^T +M_{kl}= 0\ \ (k\neq l). $$ Furthermore, by the
arbitrariness of $(r_1,\ldots, r_n)$, we must have   $M_{kl}=0$ $
(k\neq l)$,
$$A_{kj}A_{lj}^T = 0 \ (k\neq l) \eqno(2.4)$$ and
$$M_{kk}=m_kI_2, \quad A_{kj}=t_{kj} {\mathcal O}_{kj} \eqno(2.5)$$
for some real numbers $m_k, t_{kj}$ and $2\times 2$ orthogonal
matrix ${\mathcal O}_{kj}$ $j=1,2,\ldots, n$. Thus by $s_k>0$ and
Eqs.(2.4)-(2.5), for each $k$, there exists unique $j_k$ such that
$t_{kj_k}\not=0$; that is, for each $k$, all $A_{kj}=0$ except
$A_{kj_k}=t_{kj_k}{\mathcal O}_{kj_k}$. Obviously,  $j_k\not=j_l$
whenever $l\not= k$. Therefore there exists a permutation $\pi$ of $
(1,2,\ldots, n)$ so that $j_k=\pi(k)$.

\if false ?? there exists unique $k_0$ such that
$$r_{j}A_{k_0,j}A_{k_0,j}^T+M_{k_0k_0}=s_{k_0}I_2 {\rm \ for\ each\
}r_j.\eqno(2.5)$$ It follows from arbitrariness of $r_j$ that
$M_{k_0k_0}=\lambda_{k_0} I_2$ and $A_{k_0,j}=t_{k_0} \mathcal
O_{k_0}$, where $\mathcal O_{k_0}$ is a $2\times 2$ orthogonal
matrix.\fi

Write  $t_{\pi(k)}=t_{k\pi(k)}$, $\lambda_{\pi(k)}=m_k$, ${\mathcal
O}_{k\pi(k)}={\mathcal O}_{\pi(k)}$ and let $P_\pi$  be the $n\times
n$ permutation matrix corresponding to $\pi$. We can describe the
structures of $K=(A_{ij})_{n\times n}$ and $M$ as
$$K=(P_\pi\otimes I_2) (\oplus_{i=1}^n t_i \mathcal O_i) \eqno(2.6)$$
and $$M=\oplus_{i=1}^n  \lambda_{\pi(i)}I_2. \eqno(2.7)$$ As $M\geq
\pm \frac{i}{2}(\Delta - K\Delta K^T)$, one gets $\lambda_i I_2\geq
\pm\frac{1}{2} (\left(\begin{array}{cc} 0 & 1 \\ -1 &
0\end{array}\right) -t_i^2 {\mathcal O}_i\left(\begin{array}{cc} 0 &
1 \\ -1 & 0\end{array}\right){\mathcal O}^T_i)$, and this holds if
and only if $\lambda_i\geq \frac{1}{2}|t_i^2-1|$ when ${\mathcal
O}_i\left(\begin{array}{cc} 0 & 1 \\ -1 &
0\end{array}\right){\mathcal O}^T_i=\left(\begin{array}{cc} 0 & 1
\\ -1 & 0\end{array}\right)$ and $\lambda_i\geq \frac{1}{2}(t_i^2+1)$ when ${\mathcal
O}_i\left(\begin{array}{cc} 0 & 1 \\ -1 &
0\end{array}\right){\mathcal O}^T_i=\left(\begin{array}{cc} 0 & -1
\\ 1 & 0\end{array}\right)$.

Conversely, assume that $\Phi=\Phi(K,M,\bar{d})$ is a Gaussian
channel with $K=(P_\pi\otimes I_2) (\oplus_{i=1}^n t_i \mathcal
O_i)$,
 $M=\oplus_{i=1}^n  \lambda_{\pi(i)}I_2$ and $\bar{d}=0$, where  $\pi$ is a permutation of $(1,2,\ldots, n)$, $\mathcal O_i$s are some $2\times 2$ orthogonal
matrices and  $\lambda_i\geq \frac{1}{2}|t_i^2\mp 1|$ with taking
$-$ or $+$ depending on ${\mathcal O}_i$ is symplectic or not, then
$\Phi=\otimes_{i=1}^n \Phi_{i\to\pi(i)}(t_{\pi(i)}{\mathcal
O}_{\pi(i)}, \lambda_{\pi(i)}I_2,0)$ which  clearly sends every
$n$-mode Gaussian incoherent state to an $n$-mode Gaussian
incoherent, completing the proof. \hfill$\square$

{\bf Remark.} By the proof of lemma 2.1, it is ease to get a
characterization of the  a Gaussian channel $\Phi=\Phi(K,M,\bar d)$
sends every product Gaussian state into a product Gaussian state if
and only if there exists a a permutation $\pi$ on $\{1,2, ..., n\}$
such that $K=(P_\pi\otimes I_2) (\oplus_{i=1}^n K_i)$,
 $M=\oplus_{i=1}^n  M_{\pi(i)}$, where $P_\pi$ is the $n\times n$ permutation matrix
corresponding to $\pi$, and in turn, if and only if
$\Phi=\otimes_{i=1}^n \Phi_{i\mapsto \pi(i)}$, where, for each $i$,
$\Phi_{i\mapsto \pi(i)}$ is a Gaussian channel from the $i$th mode
to the $\pi(i)$th mode. In fact, if $K,M$ have the mentioned form,
then, for arbitrary $n$-mode product Gaussian state
$\rho=\otimes_{i=1}^n\rho_i$ with its covariance matrix
$\oplus_{i=1}^n\nu_i$, the covariance matrix of $\Phi(\rho)$ is
$$K(\oplus_{i=1}^n\nu_i)K^T+M=\oplus_{i=1}^n (K_{\pi(i)}\nu_iK^T_{\pi(i)}+M_{\pi(i)}),\eqno(2.8)$$
which implies that $\Phi(\rho) $ is a product state.

Next we give a characterization of $n$-mode GCBCs which reveals that
a Gaussian channel is coherence-breaking if and only if it collapses
to a Gaussian incoherent state.

{\bf Theorem 2.3} {\it An $n$-mode Gaussian channel $\Phi(K, M,
\bar{d})$ is coherence-breaking if and only if $K=0$, $\bar{d}=0$
and there exist scalars $\lambda_i\geq \frac{1}{2}$ such that
$$M={\rm diag} (\lambda_1 I_2, \lambda_2 I_2, \ldots, \lambda_n
I_2).$$}

{\bf Proof. } The ``if" part is obvious, let us check   the ``only
if" part. If an $n$-mode Gaussian channel $\Phi=\Phi(K, M, \bar{d})$
is coherence-breaking, then $\Phi$ is incoherent. By Lemma 2.1 and
Corollary 2.2, $\Phi=\otimes_{i=1}^n \Phi_{i\mapsto \pi(i)}$, where
$\Phi_{i\mapsto \pi(i)}=\Phi_{i\mapsto \pi(i)}(K_{\pi(i)},
M_{\pi(i)}, 0)$ with
$$K_i=t_i \mathcal O_i, \ \ M_i=\left(\begin{array} {ccccccccc} \lambda_i &
 0 \\
 0 & \lambda_i\\
 \end{array}\right) \eqno(2.9)$$
for some  2$\times$2 real orthogonal matrices $\mathcal O_i$ and
real numbers $t_i$ and $\lambda_i\geq \frac{1}{2}|t_i^2\mp 1|$. So,
to complete the proof, it is enough to check that each $t_i=0$.

It is clear that each $\Phi_{i\mapsto \pi(i)}$ is Gaussian coherence
breaking. Assume on the contrary  $t_i\not=0$ for some $i$. As we
know, a 2$\times$2 real orthogonal matrix has   one of the following
forms:
$$\left(\begin{array} {ccccccccc} \cos \theta &
 \sin \theta \\
 -\sin \theta & \cos \theta\\
 \end{array}\right),\quad\left(\begin{array} {ccccccccc} \cos \theta &
 \sin \theta \\
 \sin \theta & -\cos \theta\\
 \end{array}\right). \eqno(2.10)$$
For a Gaussian state $\rho$ in $\pi^{-1}(i)$th mode, suppose that
its CM is
$$\nu_\rho=\left(\begin{array} {ccccccccc} a &
 c \\
  c & b\\
 \end{array}\right),$$ where $a\geq 0, b\geq 0$, and $ab\geq c^2+\frac{1}{4}$.

 If $\mathcal O_i$ has the first form in (2.10),  then $t_i \mathcal O_i \nu_\rho \mathcal O_i^T + M_i$
 has to be diagonal for any $\rho$. It follows from a short
 computation that $$t_i[-a \cos \theta \sin \theta - c \sin^2 \theta+c \cos^2 \theta+ b\cos \theta \sin
 \theta]=0 \eqno (2.11)$$
 holds for all real numbers $a,b,c$ with $a\geq 0, b\geq 0$ and $ab\geq c^2+\frac{1}{4}$. As $t_i\neq 0$ in Eq (2.11), taking $c\neq 0$ and $a= b$ leads to $\cos^2 \theta=\sin^2
 \theta$, that is, $\cos \theta=\pm \sin \theta$. Thus $(b-a)\cos \theta \sin
 \theta$ is always  zero in Eq (2.11). Taking $a\neq b$, we get either $\cos \theta=0$ or
 $\sin\theta=0$. It follows from $\cos \theta=\pm \sin \theta$ that $\cos\theta=\sin\theta=0$, a contradiction.

Similarly, if $\mathcal O_i$ has the second form, one can also get a
contradiction.

In summary, $t_i=0$ for all $i$ and hence $K=0$. We complete the
proof. \hfill$\square$

Now let us turn to considering the relationship between some known
classes of Gaussian channels. A picture about inclusion relations
between GCBCs, GPPTCs, GEBCs, GCQCs and GQCCs will be  shown.
Without loss of generality,  in the remained part of the section, we
assume that all Gaussian channels involved have zero displacement
vectors.

We begin by recalling   the definitions of several kinds of Gaussian
channel.

 Let $H_A, H_B, H_E$ be separable
infinite-dimensional Hilbert spaces.  A Gaussian channel $\Phi:
\mathcal T(H_A)\rightarrow \mathcal T(H_B)$ is called a Gaussian
entanglement-breaking channel (GEBC) if $\Phi\otimes I(\rho^{AE})$
is always separable for any Gaussian input state $\rho^{AE}\in
\mathcal T(H_A\otimes H_E)$; $\Phi$ is called a Gaussian positive
partial transpose channel (GPPTC) if $\Phi\otimes I(\rho^{AE})$ has
the positive partial transpose for any Gaussian input state
$\rho^{AE}$  (\cite{HS}); $\Phi$ is called a Gaussian
classical-quantum channel (GCQC) if $\Phi(\rho)=\int_X \langle
x|\rho|x\rangle \rho_x {\rm d}x$, where ${\rm d}x$ is the Lebesgue
measure and $\{|x\rangle: x\in X\}$ is the Dirac's system satisfying
$\langle x|y\rangle=\delta (x-y)$, which is a direct analogy of CQ
maps in the finite-dimensional systems(Ref. \cite{Hel2}, \cite{Hel3}).

We denote by $\Omega^G_{\rm PPT}, \Omega^G_{\rm EB}, \Omega^G_{\rm
CQ}$ and $ \Omega^G_{\rm CB}$ the sets of GPPTCs, GEBCs, GCQCs and
GCBCs respetively.

To uncover the inclusion relation between the above kinds of
Gaussian channels, let us start from a general Gaussian channels.
The matrix $M$ for an $n$-mode Gaussian channel  $\Phi(K,M)$
satisfies
$$M\geq \pm \frac{i}{2}(\Delta-K \Delta K^T),$$
where $\Delta=\oplus^n_{i=1} \Delta_i$ with
$\Delta_i=\begin{pmatrix} 0 & 1 \cr -1 & 0
\end{pmatrix}$. Notice also that a Gaussian channel $\Phi(K, M)$
is entanglement-breaking (GEBC) if and only if there exist matrices
$M_1, M_2$ such that (Theorem 12.35 in Ref. \cite{Hel5}) $$M= M_1
+M_2,\ M_1\geq \pm\frac{i}{2}\Delta, \ M_2\geq \pm\frac{i}{2} K
\Delta K^T.
$$ Moreover, $\Phi$ is a GPPTC if and only if $M\geq
\frac{i}{2}(\Delta\pm  K \Delta K^T)$ (Ref. \cite{Hel5}). Then, it
follows that $$\Omega^G_{\rm EB}\subseteq \Omega^G_{\rm PPT}.$$
 Holevo (\cite{Hel2, Hel3}) showed  that $\Phi(K, M,
\bar{d})$ is a GCQC if and only if $K\Delta K^T=0$.  A direct
observation and Theorem 2.3 reveal that
$$\Omega^G_{\rm CB}\subseteq \Omega^G_{\rm
CQ}.$$

Next let us check that $\Omega^G_{\rm CQ}\subseteq \Omega^G_{\rm
EB}$. If $M$ satisfies that $M\geq \pm\frac{i}{2}K \Delta K^T$,
taking $M_1=M- (\pm\frac{i}{2}K \Delta K^T)$ and
$M_2=\pm\frac{i}{2}K \Delta K^T$. It follows that $M_1\geq 0$. So
$\Omega^G_{\rm CQ}\subseteq \Omega^G_{\rm EB}$.

In summary,   we have

{\bf Proposition 2.4.} {\it The following inclusion relations are
true.}
$$\Omega^G_{\rm CB}\subseteq \Omega^G_{\rm
CQ}\subseteq\Omega^G_{\rm EB}\subseteq \Omega^G_{\rm PPT}.
\eqno(2.12)$$

\section{Property of GCBCs with assistant entanglement inputs}

Next we are interested in the Gaussian coherence breaking channels
with assistant entanglement inputs. Such property is helpful to
understand relative problems on channel capacities
(\cite{Hel2}-\cite{Pil}). Without loss of generality, we assume that
the Gaussian channels in this section are those with zero
displacement. We first present an observation as follows.

{\bf Theorem 3.1} {\it For any Gaussian channel $\Psi$ on a system
$H_E$ and any $n$-mode Gaussian coherence breaking channel $\Phi$ on
a   system $H_A$, $\Phi\otimes \Psi(\rho^{AE})$ is always a product
state for any input Gaussian state $\rho^{AE}$ on the composite
system $H_A\otimes H_E$.}

{\bf Proof. } Suppose $\Psi=\Psi(X_\Psi, Y_\Psi)$ and $\rho_{AE}$
has the CM $\begin{pmatrix} A & C \cr C^T & E \cr
\end{pmatrix}$. Then the output $\Phi\otimes
\Psi(\rho^{AE})$ has the  CM $v_{\rm out}$ of the form
$$ \nu_{\rm out}=\begin{pmatrix}
0 & 0 \cr 0 & X_\Psi \cr
\end{pmatrix}
\begin{pmatrix} A & C \cr C^T & E \cr
\end{pmatrix}\begin{pmatrix} 0 & 0 \cr 0 & X_\Psi^T \cr
\end{pmatrix} +\begin{pmatrix}  \oplus_{i=1}^n \lambda_i I_2 & 0 \cr 0 & Y_\Psi \cr
\end{pmatrix},
$$
where $\lambda_i$ is the eigenvalue of thermal state in the $i$th
mode of the system $A$. It follows that
$$ \nu_{\rm out}=  (\oplus_{i=1}^n \omega_i I_2) \oplus (X_{\Psi} E  X_{\Psi}^T +Y_\Psi). $$
Thus, $\Phi\otimes \Psi(\rho^{AE})=\Phi(\rho_A)\otimes \Psi(\rho_E)$
is a product state.  \hfill$\square$

The observation says that the set of  GCBCs is a proper subset of
the set of  GEBCs. In order to link to applications in the theory of
channel capacities, it is needed to discuss the property of the
$k$-tensor-product for GCBCs and arbitrary channels with entangled
inputs.

Suppose that the Gaussian channel $\Psi$ has the form $\Psi(X_\Psi,
Y_\Psi)$ with the zero displacement vector and $\Phi\otimes \Psi$
acts on bipartite system $H_A\otimes H_B$. It follows that the
Gaussian channel $(\Phi\otimes \Psi)^{\otimes k} $ can be described
based on covariance matrices of input states as follows
$$ \nu_{\rm in}\mapsto\begin{pmatrix}
0 & 0 \cr 0 & X_\Psi \cr
\end{pmatrix}^{\oplus k}
\nu_{\rm in}\begin{pmatrix} 0 & 0 \cr 0 & X_\Psi^T \cr
\end{pmatrix}^{\oplus k} +\begin{pmatrix}  \oplus_{i=1}^n \omega_i I_2 & 0 \cr 0 & Y_\Psi \cr
\end{pmatrix}^{\oplus k},
$$
where diag\ $\nu_{\rm in}={\rm diag}(\nu_{AB}, \nu_{AB},...,
\nu_{AB})$, $\nu_{AB}=\begin{pmatrix} A & C \cr C^T & B \cr
\end{pmatrix}$
is the covariance matrix for input states of $\Phi\otimes \Psi$,
$\omega_i$ is the eigenvalue of thermal state in the $i$th mode of
the system $A$. It follows that the CM $\nu_{\rm out}$ of the output
state
$$ \nu_{\rm out}=\begin{pmatrix}  \oplus_{i=1}^n \omega_i I_2 & 0 \cr 0 & X_{\Psi} B  X_{\Psi}^T +Y_\Psi  \cr
\end{pmatrix}^{\oplus k}. $$
Without loss of generality, write $\nu_{\rm mod}=(m_{ij})$ the
modulation covariance matrix for the Gaussian channel $(\Phi\otimes
\Psi)^{\otimes k}$  (for example, see \cite{Sch}, \cite{Sch2}).
Similar to the above discussion, we have that the modulated output
state has the covariance matrix $\bar{\nu}_{\rm out}$ of the form
$$ \bar{\nu}_{\rm
out}=\nu_{(\Phi\otimes \Psi)^{\otimes k}(\rho[\nu_{\rm in}+\nu_{\rm
mod}])}=
\begin{pmatrix} \oplus_{i=1}^n \omega_i I_2 & 0 \cr 0 &
X_{\Psi} (B+\nu'_{\rm mod}) X_{\Psi}^T +Y_\Psi \cr
\end{pmatrix}^{\oplus k}, \eqno(3.1) $$
where   $\nu'_{\rm mod}$ is the submatrix of $\nu_{\rm mod}$
corresponding to $B$.

Recall that the capacity of quantum channels is a core topic in
quantum information theory, and the additivity question for channel
capacity arises from the following one: whether or not the entangled
inputs can improve the classical capacity of quantum channels
(\cite{Shor}-\cite{Pil}). The classical $\chi$-capacity of a channel
$\Phi$ is defined as
$$C_\chi (\Phi)= \sup_{\{p_i, \rho_i\}} [S(\sum_i p_i(\Phi(\rho_i)))-\sum_i p_i S(\Phi(\rho_i)], \eqno (3.2)$$
where $S$ denotes the quantum entropy. The classical capacity
$C(\Phi)$ is defined by
$$C(\Phi)= \lim_{k\rightarrow\infty}\frac{1}{k} C_\chi (\Phi^{\otimes k}). \eqno (3.3)$$
If $\Phi$ is a Gaussian channel, its Gaussian classical
$\chi$-capacity is defined as
$$C_\chi^G (\Phi)= \sup_{\mu, \bar{\rho}} S(\Phi(\bar{\rho}))-\int \mu({\rm d} \omega) S(\Phi(\rho(\omega))),  \eqno (3.4)$$
where the input $\rho(\omega)$ is $n$-mode Gaussian state,
$\bar{\rho}= \int \mu ({\rm d}z)\rho(z)$ is the so-called averaged
signal state, $\mu$ is a probability measure.  Note that the right
side of Eq (3.4) may be infinite. It is obvious that the Gaussian
classical capacity is a lower bound of the classical capacity. The
additivity of channel capacity can be described as
$$\mathcal C(\Phi\otimes \Psi)=\mathcal C(\Phi)+\mathcal C(\Psi),\eqno(3.5)$$
where $\Phi$ and $\Psi$ are channels and $\mathcal C$ denotes the
channel capacity (\cite{Ben}). The additivity of classical capacity
for  GEBCs had been discussed in (\cite{Hel6}, \cite{Shi}). Here, we
are interested in giving a shorter proof of the additivity of
classical capacity of the tensor product of a GCBC with   a Gaussian
channel. For the Gaussian channel $\Psi$ of the form $\Psi(X_\Psi,
Y_\Psi)$ and the GCBC $\Phi$, it follows from (3.1) that
$$\begin{array}{llll} &C_\chi((\Phi\otimes \Psi)^{\otimes k}) \\& =S(\rho[\bar{\nu}_{\rm out}])-S(\rho[\nu_{\rm
out}])\\ &= kS(\rho[\oplus_{i=1}^n \omega_i I_2]\otimes
\rho[X_{\Psi} (B+\nu'_{\rm mod}) X_{\Psi}^T +Y_\Psi])-\\ &\ \ \ \
kS(\rho[\oplus_{i=1}^n
\omega_i I_2]\otimes \rho[X_{\Psi} B X_{\Psi}^T +Y_\Psi]) \\
&=kS(\Phi(\bar{\rho}_A))+kS(\Psi(\bar{\rho}_E))-kS(\Phi(\rho_A))-kS(\Psi(\rho_E))
\\ &= kC_\chi(\Phi)+kC_\chi(\Psi).
\end{array} $$
It follows   that the Gaussian classical capacity and Gaussian
classical $\chi$-capacity are both additive for the tensor product
of a GCBC and arbitrary a Gaussian channel.

\section{Conclusion}

(Need rewrite) We have give  a complete characterization of
arbitrary $n$-mode Gaussian coherence breaking channels (GCBCs),
which constitute a proper subclass of Gaussian entanglement breaking
channels. Indeed, Gaussian coherence breaking channels have more
rigorous properties, comparing with Gaussian entanglement breaking
ones. A obvious witness is our observation that the tensor product
of a GCBC and arbitrary a Gaussian channel maps all input states
into product states. We also establish the inclusion relations among
GCBCs and other common kinds of Gaussian channels. Furthermore, we
discuss the property of the $k$-tensor-product of GCBCs and
arbitrary channels with entangled inputs, and show a simple
application in the research on the theory of the channel capacity.

{\bf Acknowledgements}  Thanks for comments. The work is supported
by National Science Foundation of China under Grant No. 11771011 and
Natural Science Foundation of Shanxi Province under Grant No.
201701D221011.

\end{document}